\documentclass[12pt, english]{article}

\usepackage[utf8]{inputenc}
\usepackage[T1]{fontenc}
\usepackage{babel}
\usepackage{amsmath}
\usepackage{graphicx}
\usepackage{fancyhdr}
\usepackage{hyperref}
\usepackage[margin=1.6in]{geometry}
\usepackage{epstopdf}
\usepackage{xcolor}
\begin{document}

\title{Multiple Regions of Shock-accelerated Particles during a Solar Coronal Mass Ejection}

\author{Diana E. Morosan\textsuperscript{1,2,*}, Eoin P. Carley\textsuperscript{1,3}, Laura A. Hayes\textsuperscript{1,3}, \\
Sophie A. Murray\textsuperscript{1,3},  Pietro Zucca\textsuperscript{4}, Richard A. Fallows\textsuperscript{4}, \\
Joe McCauley\textsuperscript{1}, Emilia K. J. Kilpua\textsuperscript{2}, Gottfried Mann\textsuperscript{5}, \\
Christian Vocks\textsuperscript{5}, Peter T. Gallagher\textsuperscript{1,3}
}
\maketitle
\thispagestyle{fancy}

\noindent
1. School of Physics, Trinity College Dublin, Dublin 2, Ireland. \\ 
2. Department of Physics, University of Helsinki, P.O. Box 64, Helsinki, Finland. \\
3. School of Cosmic Physics, Dublin Institute for Advanced Studies, Dublin, D02 XF85, Ireland. \\
4. ASTRON, Netherlands Institute for Radio Astronomy, Oude Hoogeveensedijk 4, 7991 PD, Dwingeloo, The Netherlands. \\
5. Leibniz-Institut f\"{u}r Astrophysik Potsdam (AIP), An der Sternwarte 16, 14482 Potsdam, Germany. \\
*morosand@tcd.ie \newline

\newpage 

\textbf{The Sun is an active star that can launch large eruptions of magnetised plasma into the heliosphere, called coronal mass ejections (CMEs). These ejections can drive shocks that accelerate particles to high energies, often resulting in radio emission at low frequencies (<200 MHz). To date, the relationship between the expansion of CMEs, shocks and particle acceleration is not well understood, partly due to the lack of radio imaging at low frequencies during the onset of shock-producing CMEs. Here, we report multi-instrument radio, white-light and ultraviolet imaging of the second largest flare in Solar Cycle 24 (2008--present) and its associated fast CME (3038$\pm$288 km/s). We identify the location of a multitude of radio shock signatures, called herringbones, and find evidence for shock accelerated electron beams at multiple locations along the expanding CME. These observations support theories of non-uniform, rippled shock fronts driven by an expanding CME in the solar corona.}

%%%%%%%%%%%%%%%%%%%%%%%%%%%%%%%%%%

%{The acceleration of charged particles to high energies is widespread throughout the Universe and it is a major topic of study in both space and laboratory plasmas.}

{Particles accelerated in collisionless shocks are of particular interest in space plasmas and are often associated with CMEs from the Sun. Shocks and related high-energy particles can propagate through the heliosphere, influencing planetary ionospheres and atmospheres, and also affecting technological systems at Earth (for a review see \cite{kr12}). Such processes are not limited to our solar system; other stars are expected to produce even larger CMEs, stronger shocks and more powerful particle acceleration \cite{od17}. Particles accelerated by these powerful eruptions from other stars can even affect the habitability of exoplanets \cite{ai17}. Since observations of stellar eruptions are very limited, studying particle acceleration at the Sun is of crucial importance for understanding these processes universally. }

{Fast CMEs (with speeds up to $\sim$3,500 km/s \cite{yu05,ru11}) are powerful drivers of plasma shocks that can accelerate particles up to relativistic speeds producing bursts of plasma emission at radio wavelengths \cite{kl02}. The most obvious manifestations of shocks at radio wavelengths on the Sun are a class of radio bursts, Type II bursts, mostly observed at frequencies <150~MHz \cite{ma96,ne85, wi50}. They usually show two emission lanes slowly drifting to lower frequencies in dynamic spectra, with a 2:1 frequency ratio representing emission at the fundamental and harmonic plasma frequency. Type II bursts have been imaged on multiple occasions showing sources closely associated with CMEs \cite{ne85, st74, sm70}, while simulations and CME reconstructions closely associate Type IIs with CME shocks \cite{schmidt13, zu18}. In some cases, `bursty' signatures of individual electron beams accelerated by CME shocks can be identified in dynamic spectra superimposed on Type II bursts \cite{ma05}. These electron beam signatures, called `herringbones', are identified as narrow bursts of radiation drifting towards higher and lower frequencies, categorised as distinct emission from the accompanying Type II burst \cite{ca87, ca89}, and sometimes even observed without a Type II \cite{ca87, ku65}. Despite the wealth of publications on Type II bursts, there has been no direct imaging of herringbones and, therefore, no direct association of individual electron beams with coronal shocks.}

% and lasting up to 10 minutes 

{Herringbones can show where particles are accelerated and where shocks occur in the corona, providing important diagnostics of shock-plasma physics. The observational characteristics of herringbones can be used to study the shock particle acceleration mechanism, and their kinematic properties can provide clues into the magnetic field geometry encountered by the shock as the accelerated electrons are believed to escape along open magnetic field lines near the shock front \cite{ho83}. However, there is still much debate regarding the origin of herringbones and two main theories have been proposed. The first theory suggests that particles are accelerated in a magnetic trap inside the throughs of a `wavy' shock \cite{zlobec93, vandas11, carley15} to explain the bursty nature of this emission. The acceleration is believed to occur at the flank of a CME \cite{schmidt12, ca13}. The second theory suggests that herringbones may be generated by a termination shock from magnetic reconnection outflow jets \cite{aurass02, mann09}. Imaging of shock signatures at low radio frequencies is thus crucial in determining the particle acceleration mechanism and also the properties and location of the shock with respect to the CME.} 

{In this study, we present evidence for multiple locations of shock accelerated particles traveling with the expanding CME. We also provide observational evidence of radio signatures of individual electron beams that are indeed accelerated by a shock at the CME flanks. }

\section{Multi-wavelength Observations}

{On 10 September 2017, the second largest flare of Solar Cycle 24 erupted, classified as an X8.2 class flare. The flare was associated with a very fast CME (3038$\pm$288 km/s), observed at EUV wavelengths with the Atmospheric Imaging Assembly (AIA; Figure 1a) \cite{le12} and the Solar Ultraviolet Imager (SUVI; Figure 1b) \cite{se18} and white-light coronagraph images from the Large Angle Spectrographic Coronagraph (LASCO; Figure 1b) \cite{br95}. Both flare and CME occurred on the western solar limb, when imaged from the Earth's perspective, providing a side view of the eruption. The CME eruption can be seen in Figure 1a-b starting with coronal loops rising, the formation of a plasmoid and the sudden release of radiation (brightening) in the low corona (Figure 1a), followed by the expansion of the CME plasmoid through the high corona (Figure 1b). A clear CME-driven shock is also visible in the middle panel of Figure 1b, identified as the sharp outer boundary of the CME plasmoid \cite{vo13}. Shock signatures were identified at radio wavelengths with the Low Frequency Array (LOFAR) \cite{lofar13}. The LOFAR core was used in tied-array beams mode (see Methods) to image herringbones with high temporal (10~ms) and spectral (12.5~kHz) resolution. Complementary dynamic spectra were also obtained using the Irish LOFAR station (I-LOFAR; Figure 1c). High-resolution zoomed-in regions are shown in the panels of Figure 1d observed with the LOFAR core.}

{The majority of the radio emission observed consists of herringbones, but many other radio features are also present (Figure 1c-d). We identified three groups of herringbones, that are divided by their separate spatial locations discussed in the next section. The first group of herringbones (Group 1) consists of the brightest shock signatures observed at low radio frequencies during the flare, starting at 15:59:10~UT in the range 30--50~MHz (first panel of Figure 1d). These herringbones have a duration of up to 2~s and electron beam speeds of $\sim$0.2~c, estimated from their drift rate (for more details see Methods). The high speed suggests the presence of non-relativistic electron beams accelerated during the flare. The second group of herringbones (Group 2; middle panel of Figure 1d) also occurs at frequencies of 30--50~MHz after Group 1, but they have shorter durations of up to 1.5~s. Furthermore, they occur more frequently and in larger numbers than Group 1, with estimated electron beam speeds of $\sim$0.25~c. The third group of herringbones (Group 3; right panels of Figure 1d) represents the fine structure of a Type II burst starting at 16:04~UT. The Type II burst consists of two emission lanes with $\sim$0 frequency drift, at the fundamental (30~MHz) and harmonic (60~MHz) plasma frequency (Figure 1c). The higher frequency lane (48--60~MHz), observed with the LOFAR core, consists of short duration herringbones ($\sim$1~s) with beam speeds of $\sim$0.25~c. }
 
 %{The presence of the Type II structure, with both the fundamental and harmonic components, indicates a different emission location to Groups 1 and 2.}

\section{Radio imaging}

{The LOFAR tied-array beam observations of the Sun offer the unique possibility of imaging herringbones at any observed frequency, with high temporal resolution \cite{mor14, mo15}. The herringbone source locations are shown in Figure 2, where filled contours at 35, 45 and 50~MHz, respectively, are overlaid on AIA 211~\AA~ (the first two and last panels of Figure 2) and SUVI 193~\AA~ (all remaining panels of Figure 2) running difference images, that show the CME expansion. The radio sources imaged in Figure 2 and throughout the paper have been corrected for ionospheric refraction (for more details see Methods). The top three panels of Figure 2 show the locations of Group 1 herringbones, the middle panels show the Group 2 herringbones and the bottom panels show the Group 3 herringbones. The filled contours represent the 70\% to maximum intensity levels in each image. The number of contours is a measure of brightness relative to the most intense radio source in all of the images (the more contours, the brighter).}

{The CME is best observed in the upper right panel of Figure 2 at 15:59:34~UT, after which it expands outside the SUVI field-of-view. As the CME expands, it is accompanied by a coronal EUV wave in the low corona, also known as a coronal bright front. The EUV wave represents a disturbance of the low coronal plasma caused by the passage of the CME and the location of the EUV wave in the low corona can be used as an approximate location of where the CME flank is projected upwards in the high corona \cite{lo08}. Figure 2 consists of still images from the Supplementary Movie accompanying the paper, which shows the three herringbone groups locations and the coincident movement of Group 1 and 2 in the direction of the EUV wave. Groups 1 and 2 travel towards the Sun in the plane-of-sky images, however, this is a projection effect. These sources move in the direction of the EUV wave at higher altitudes, following the expansion of the southern CME flank to the left. Group 3 is located in the north and it moves in the direction of the northern CME flank.}

{There is a clear white-light shock at both the northern and southern CME flanks (denoted by arrows in Figure 1b) indicating that the herringbone sources, also located at the flank, are accelerated at the shock front. The 3D representation of the spheroidal shock bubble surrounding the CME volume is overlaid in the last panel of Figure 2, which extends from the high corona to the low corona where the EUV wave is observed as a CME-driven disturbance. Groups 1 and 2 are located on top of this shock bubble in 3D, at the southern CME flank, while Group 3 is consistent with the shock location at the northern CME flank. The association of CMEs, coronal EUV waves and shock particle acceleration has been observed before in the case of Type II radio bursts \cite{ca13, gr11}, however, individual herringbones, have not been directly related to these phenomena until now. The synchronised movement of the EUV wave, CME flank and herringbones sources indicates that individual electron beams are accelerated by a shock at the CME flank and not at a termination shock region.} 
%{These findings represent the best observational evidence to date that show individual electron beams accelerated by shocks at the flanks of CMEs. }

\section{Radio source propagation}

{Herringbone electrons escape along open magnetic field lines after reflecting from the shock \cite{ca13}, therefore, radio imaging can be used to determine the geometry of the open field. Herringbones from Group 2 experience a frequency separation in space that changes with time (Figure 3). The 35 MHz source, in the plane-of-sky, is initially located further away from the western limb than the 50 MHz emission at $\sim$16:00~UT (left panel of Figure 3a). After $\sim$2~minutes, as the source moves to the left in the direction of the coronal wave, the 35~MHz emission appears closer to the western limb than the 50~MHz source (left panel of Figure 3b). Based on the herringbone frequency drift in Figure 1b, the lower frequency sources in Figure 3 (35 and 40~MHz) are traveling away from the Sun, since they drift towards lower frequencies (forward drift), and the higher frequency sources (45 and 50~MHz) are traveling towards the Sun, as they drift towards higher frequencies (reverse drift). This occurs due to the plasma emission mechanism where the frequency of emission $f \approx 8980 \sqrt{n_e}$, where $f$ is in Hz and $n_e$ is the electron density in cm$^{-3}$, which decreases radially from the Sun. }

{A possible scenario of the magnetic field geometry at both times is shown in the cartoons in Figure 3 from a solar north perspective (middle) and the same line-of-sight perspective as in the left panels (right). Since the accelerated particles reflected at the shock escape along open field lines, this frequency orientation can be explained by the CME either deflecting open magnetic field lines in its path or encountering new, bent field lines (Figure 3b). }

%The frequency separation observed, may also be affected by radio waves propagation effects in the corona, at low frequencies \cite{}.}

{The herringbone radio sources are therefore located away from the sky-plane towards the observer, based on the cartoon in Figure 3. In the plane-of-sky, an individual herringbone from Group 2, with forward and reverse drift components (Figure 4a), is located at the southern CME flank, on top of the shock bubble surrounding the CME (Figure 4b). The higher frequency sources are located closer to the shock bubble (orange dots) travelling towards the Sun (reverse drift frequencies) and the lower frequencies further away travelling away from the Sun (forward drift frequencies). The higher frequency sources appear to be located on the line-of-sight moving away from the observer. }

{There is a clear shock at the southern and northern flank in the plane-of-sky (Figure 1b), however, it is necessary to determine if the CME also drives a shock away from the plane-of-sky, in a direction towards the observer. Therefore, there is a need to determine if the radio sources and CME flank are located in a region of low Alfv\'en speed in three-dimensional space, for the CME to drive a shock. This requires the construction of a 3D space of Alfv\'en speed in a volume through which the CME erupts using electron density maps \cite{zu14} and extrapolations of the surface magnetic field (for more details, see Methods). Five Alfv\'en speed surfaces at consecutively larger radii are shown in Figure 4c, where the inset to the figure clearly shows a region of high Alfv\'en speed of $\sim$1000 km/s (red areas) above the active region. The extrapolated field lines are shown to indicate regions of closed and open field in this environment. The three dimensional representation of the spheroidal shock surface is shown as the black points in Figure 4c, representing the location of the CME shock in the 3D Alfv\'en speed environment. Lastly, we attempt to de-project the radio sources from the sky-plane using the 3D density environment constructed above. To do this, we firstly convert the frequency of each radio source to density, assuming fundamental plasma emission. Taking the x-y coordinates of the radio centroids in the sky-plane, we search for the densities for each radio source along the line of sight (in the Sun-Earth line direction) in the 3D density environment. We take the location of this density as the z-coordinate of the radio source in the 3D environment. The coloured spheres in Figure 4c show the locations of the radio sources in 3D, with frequencies as indicated. Such a de-projection cannot give the precise location of the radio sources in 3D space, but is used to give a rough indication of the general environment into which the herringbone electron beams propagate. }

{The analysis shows that the radio sources are located on the southern flank of the erupting structure in a region of relatively low Alfv\'en speed of ~200 km/s (dark blue regions) and open magnetic field towards the south (the purple field lines). There is, however, a misalignment of herringbone electrons propagation and open field towards the south in Figure 4c. This is due to inherent uncertainties in the de-projection method and the open magnetic field lines extrapolation, without taking into account the effect of the CME expansion on the open field. We caution that this should not be interpreted as electron beams crossing the field lines, the method cannot predict such precise locations of radio sources in 3D space. Rather we use this method to determine the general environment into which the beam propagates, making no claim of how the beams propagate with respect to open field. Finally, we also note that the forward drift herringbones propagate further than the reverse drift, predicted in the analysis of \cite{carley15}.}

\section{Shock Properties}

{The full extent of the CME during its early expansion stage is shown in Figure 5a in the SUVI and LASCO composite running difference images. Overlaid on these images are the centroids of the sources (filled circles) at 45~MHz (Groups 1 and 2) and 50~MHz (Group 3) colour-coded through time from 15:59:09--16:05:33~UT. The centroids motions, in the direction of the yellow arrows, confirm that Groups 1 and 2 move in the same direction as the EUV wave and Group 3 moves in the direction of the northern CME flank. We can estimate the shock Mach number at the flanks by measuring the CME expansion along the northern and southern flanks and computing the plane-of-sky Alfv\'en speed. The Alfv\'en speed map in Figure 5b shows that, in the plane-of sky, Groups 1, 2 and 3 occur at locations where the CME encounters regions of Alfv\'en speed minima. These minima are also observed off the plane-of-sky in the 3D Alfv\'en speed volume (Figure 4c). Therefore, the shock Mach number in the plane-of-sky is a close estimate for the magnitude of the shock in the 3D Alfv\'en speed minima volumes, where the radio sources are located.}

{The radial CME speed with height was estimated using plane-of-sky AIA, SUVI and LASCO images (for more information, see Methods). The CME accelerates very rapidly in the low corona and reaches a maximum speed of 3038$\pm$288~km~s$^{-1}$ at $\sim$4~R$_{\odot}$ (Figure 6a), making it one of the fastest CMEs ever observed \cite{yu05, be11, liu13}. The coloured vertical dashed lines in Figure 6a represent the heights corresponding to radio sources at frequencies of 35--50 MHz, based on the constructed density maps, indicating the approximate height where electron acceleration occurs at these frequencies. At these heights, the CME is already super-alfv\'enic compared to the two Alfv\'en speed profiles (orange and purple lines in Figure 6a taken along the white dashed lines in Figure 5b). }

{To estimate the Mach number of the shock at the flanks, we computed the CME flank speed for the southern (Figure 6b) and northern (Figure 6c) flanks, along the red dashed arcs in Figure 5b towards the southern and northern pole, respectively. The CME flank speed shows that the southern sources (Groups 1 and 2) occur in a plane-of-sky region where the CME encounters a 'dip' in the Alfv\'en speed (Figure 6b). In the case of the northern sources (Group 3), the CME becomes super-alfv\'enic at a later time in good agreement with the later appearance of Group 3  (Figure 6c). At the southern flank, inside the dip at a height of $\sim1.4~R_{\odot}$, the CME has a high flank speed of $1100\pm500$~km~s$^{-1}$, similar to values reported by \cite{we18}. At the same location, the Alfv\'en speed is low ($\sim$380~km~s$^{-1}$) and therefore the CME produces a shock of Mach 2.9. At the northern flank, at $\sim$1.7~R$_{\odot}$, the Alfv\'en speed is $\sim$500~km~s$^{-1}$ and CME speed is $\sim$840$\pm480$~km~s$^{-1}$. Here, the CME drives a shock of Mach 1.7. The CME driven shock of Mach 2.9 is a strong shock when compared to previously reported CME shocks in the low corona \cite{be13, shen2007}. We also note that the brightest radio emission occurred at the southern CME flank, where the shock Mach number was significantly higher than that of the northern CME flank.}

{In our observations, the CME was sufficiently fast to drive a shock at both the northern and southern flanks, in regions of Alfv\'en speed minima. The multitude of electron beam signatures observed, suggests that particles easily escaped the shock front, most likely due to the presence of open magnetic field lines encountered by the CME, that are indeed present at both the southern and northern flanks. The near-relativistic electron beam velocities of 0.2--0.25~c are sufficient to generate plasma instabilities such as Langmuir waves to generate the observed herringbones following wave-wave interactions. Such high speeds can be achieved in cases where the direction of the shock is quasi-perpendicular to the ambient magnetic field \cite{ho83}. Studies at low radio frequencies, have shown that the angle between the shock direction and upstream magnetic field is $>$87$^\circ$ \cite{mann2018}. The generation of multiple electron beams with upstream and downstream directions supports the theory of electrons being generated at the 'bulge' of a 'rippled' shock front \cite{ho83,zlobec93,vandas11}. Electron beam reflections from the shock then occur on either side of the bulge for upstream and downstream electrons to escape. Our observations indicate that the shock front is inhomogeneous on both the northern and southern flank of the CME. Observations of collisionless shocks at Earth confirm that surface ripples are indeed present at the shock front on small spatial and temporal scales \cite{jo16, gi17}. The small spatial and temporal scales of the ripples, if present at the Sun, can explain the short temporal width of herringbones in dynamic spectra. Simulations of electron beams accelerated at a rippled shock front in the Earth's magnetosphere, show that electrons are not reflected uniformly in time and space, but packs of reflected electrons are formed along the rippled shock front \cite{le02, ya18}. A similar mechanism at the Sun can explain the diversity of herringbones observed with slightly varying spectral characteristics (duration, bandwidth and acceleration frequency where the reverse and forward drift components meet), generated by bursty 'packs' of electron beams. }

\section*{Correspondence}
{Correspondence and requests for materials should be addressed to D.E.M.}

\section*{Acknowledgements}

This paper is based (in part) on data obtained with the International LOFAR Telescope (ILT) under project code DDT8\_005. LOFAR (van Haarlem et al. 2013) is the Low Frequency Array designed and constructed by ASTRON. It has observing, data processing, and data storage facilities in several countries, that are owned by various parties (each with their own funding sources), and that are collectively operated by the ILT foundation under a joint scientific policy. The ILT resources have benefited from the following recent major funding sources: CNRS-INSU, Observatoire de Paris and Universit\'e d'Orl\'eans, France; BMBF, MIWF-NRW, MPG, Germany; Science Foundation Ireland (SFI), Department of Business, Enterprise and Innovation (DBEI), Ireland; NWO, The Netherlands; The Science and Technology Facilities Council, UK. I-LOFAR received funding from Science Foundation Ireland (SFI) grant no. 15/RI/3204. D.E.M received external funding from the MET Office, Exeter, UK at Trinity College Dublin. E.P.C. is supported by the H2020 INFRADEV-1-2017 LOFAR4SW project no. 777442.  L.A.H. is supported by Enterprise Partnership Scheme studentship from the Irish Research Council (IRC) between Trinity College Dublin and Adnet System Inc. S.A.M. is supported by the Irish Research Council Postdoctoral Fellowship Programme and the Air Force Office of Scientific Research award number FA9550-17-1-039. E.K. and D.E.M. acknowledge The Finnish Centre of Excellence in Research of Sustainable Space, funded through the Academy of Finland grant no. 312351 and Academy of Finland Project 1310445. This project has received funding from the European Research Council (ERC) under the European Union's Horizon 2020 research and innovation programme (grant no. 4100103, SolMAG). The authors would like to acknowledge NOAA's National Centers for Environmental Information SUVI team for providing data on the September 2017 flares and Maxime Grandin for his advice in correcting for ionospheric effects.

\section*{Author Contributions}

{D.E.M. performed the data analysis, interpretation of results and prepared the manuscript. E.P.C reconstructed the 3D model, contributed to the interpretation of results and manuscript preparation. L.A.H. produced the cartoon, contributed to the preparation of EUV images and interpretation of results. S.A.M. processed the EUV images and contributed to discussion of the results and manuscript preparation. P.Z. supplied the Alfv\'en speed maps and prepared the LOFAR core observation with the help of R.A.F. J.M. prepared and supplied the I-LOFAR observations. E.K. contributed to the interpretation of results and manuscript preparation. G.M and C.V were involved in the LOFAR observing proposal. P.T.G. is the PI of the LOFAR observing proposal and I-LOFAR project and guided the data analysis and writing of the manuscript. }

\newpage

\section*{Methods}

\subsection*{EUV Imaging}
{The AIA 211~\AA~ images in Figure 1a were processed with the multi-scale Gaussian normalization (MGN) technique of \cite{mo14} to enhance the coronal loops and plasmoid eruption. Figure 2 is composed of both AIA 211~\AA~ (top two and last panel) and SUVI 193~\AA~ \cite{suvi} (all remaining panels) running difference images. The SUVI 193~\AA~ images in Figure 2 have been chosen to provide a larger field-of-view than AIA in order to show the erupting plasmoid up to higher heights. AIA images were used for the better cadence. }

\subsection*{Electron Beam Velocity Estimates}
{The frequency of plasma emission, $ f_p$, is directly proportional to the electron plasma density and it can be expressed by the following relation: $ f_p  = C \sqrt{n_e}$, where $C$ is a constant (see Equation 1) and $n_e$ is the electron density. For radio emission at the plasma frequency ($f = f_p$), the frequency drift rate $df/dt$, can be used to determine the velocity, $v$, of the electron beam generating plasma emission using the following relation:
\begin{equation}
v = \frac{2\sqrt{n_e}}{C} \Big( \frac{dn_e}{dr} \Big)^{-1} \frac{df}{dt}
\end{equation}
The unknown in the above equation is plasma density, $n_e$. Since the electron density decreases radially in the corona, it is necessary to consider a density model, $n_e = n_e(r)$, to obtain the density at a specific height. We used the Newkirk density model \cite{new61} to obtain the velocities of herringbone electrons in Section 1. The drift rate, $df/dt$, of herringbones was estimated directly from dynamic spectra by computing the ratio between total frequency bandwidth and total duration of each herringbone.}

\subsection*{LOFAR Tied-array Imaging}
{LOFAR is a new-generation radio interferometric array constructed by the Netherlands Institute for Radio Astronomy (ASTRON), operating in the frequency band of 10--240~MHz \cite{lofar13}. The I-LOFAR dynamic spectrum in Figure 1 was observed using the Low Band Antennas (LBAs) of the international LOFAR station in Ireland during an observation campaign in stand-alone mode to provide spectral characteristics of solar radio bursts. The LOFAR core dynamic spectra were recorded using the core LBA stations in the Netherlands and they show unprecedented high-time and frequency resolution observations. Unfortunately, observations with the LOFAR core ended at 16:06~UT, therefore imaging is not possible after this time.  }

{Here, we used one of LOFAR's beam formed modes, tied-array beams \cite{st11, lofar13}, at LBA frequencies of 30--70~MHz. The 24 LOFAR core stations, with a baseline of $\sim$2~km, were used to produce 217 simultaneous tied-array beams to observe the Sun covering a field-of-view of $\sim$1.5$^{\circ}$. The pattern of the 217 beams is shown in Supplementary Figure 1 where the yellow circle is the position of the Sun relative to the tied-array beam pattern at 16:00~UT. Each beam recorded a high time and frequency resolution dynamic spectrum ($\sim$10~ms; 12.5~kHz), for example Figure 1d, in which the radio bursts can be identified. Since each beam has a spatial location, the intensity of radiation at that location can be extracted from the dynamic spectra. These intensity values can be plotted as `macro-pixels' onto a tied-array map for a chosen time and frequency. Radio sources can then be imaged by interpolating between the intensity of radiation from all beams and producing tied-array images of radio bursts (for more details, see \cite{mor14, mo15}).  Images of radio bursts presented here were averaged over a frequency range of 0.5~MHz to reduce spectral noise and over a time period of 0.5~s in order to capture the full temporal extent of the herringbones. }

\subsection*{Fitting the Shock Bubble}
{To clearly show the location of the herringbone sources above the 3D CME volume, we perform a fit by eye of the erupting volume as imaged in AIA and SUVI images. To do this, we orient a 3D wireframe sphere until it matches best with the EUV wave propagating north and south and onto the solar surface. The EUV wave (spheroidal surface) is assumed to be the low coronal manifestation of the shock driven by the erupting CME. The spheroidal surface therefore extends from the high corona to the low corona, up to the locations of the EUV wave. }

\subsection*{Alfv\'en Speed Maps}
{The Alfv\'en speed maps were estimated based on the methods of \cite{zu14} for 10 September 2017. Electron densities in the corona are estimated in two ways: for the height range 1--1.3~R$_\odot$, densities were obtained from the differential emission measure derived from the six coronal filters of AIA and, for the height range 2.5--5~R$_\odot$ they were obtained using polarised brightness from the Large Angle and Spectrometric Coronagraph (LASCO) \cite{br95} on-board the Solar and Heliospheric Observatory (SOHO). For the intermediate height range 1.3--2.5~R$_{\odot}$, a combined plane-parallel and spherically symmetric model was employed (see \cite{zu14} for further details). Using these electron densities we estimated the source heights at frequencies of 35 to 50~MHz denoted by the coloured dashed lines in Figure 3c. The magnetic field was estimated using a potential field source surface (PFSS) \cite{sc03} model that provides an approximation of the coronal magnetic field at heights up to 2.5~R$_{\odot}$ based on the observed photospheric field. We used the PFSS solution from 10 September 2017 when X8.2 flare active region was on the limb. An example of magnetic field lines extrapolated using PFSS is shown in Supplementary Figure 2, where an extensive area of open field lines can be seen at the western solar limb before (left) and after (right) the CME occurred on 10 September 2017. Bent or deflected open magnetic field lines were also observed in a south-westerly direction. } 

{The Alfv\'en speed in a plasma is then calculated using the following relation:
\begin{equation}
v_A = \frac{B}{\sqrt{\mu_0 n_i m_i}}~~,
\end{equation}
where $v_A$ is the Alfv\'en speed, $B$ is the magnetic field strength, $n_i$ is the ion density, $m_i$ the ion mass and $\mu_0$ is a known physical constant. }

{Plane-of-sky images of Alfv\'en speed can be constructed by combining the density map from 10 September 2017 and the plane-of-sky slice of the PFSS solution. To construct the 3D Alfv\'en speed environment, we use the plane-of-sky density map and assume a rotational symmetry of this density map about the north solar axis, giving us a 3D volume of densities above the western solar hemisphere. We then use magnetic field extrapolations of the photospheric field to determine the magnetic field strength at each point in the same 3D volume. This allows us to construct a 3D Alfv\'en speed environment above the eruption site in question. }

\subsection*{Fitting the CME Velocity Function}
{The CME velocity fit was estimated by fitting the height-time function, $h(t)$ from \cite{ga03}, to plane-of-sky AIA, SUVI and LASCO images following the expanding CME front radially in the positive x-direction:
\begin{equation}
h(t) = h_0 + v_0t + \int\int \Big(\frac{1}{a_r e^{t/\tau_r}} + \frac{1}{a_d e^{-t/\tau_d}} \Big)^{-1} dt dt ,
\end{equation}
where $h_0, v_0$ are the initial heights and velocities, respectively, and the term inside the double integral represents the CME acceleration function with time, where $a_r, a_d$ are the initial accelerations and $\tau_r, \tau_d$ are the e-folding times for rise and decay phases of the acceleration. The 10 September 2017 CME fit can be seen in Supplementary Figure 3 where the first panel shows the CME height in time, the second panel shows the CME speed and the third panel shows the CME acceleration. Since the CME is located at the limb, the speed estimate of the CME following the fitting procedure is expected to be accurate. The CME accelerates very fast within the first 2 minutes and then reaches a top speed of 3000$\pm$300~km s$^{-1}$ (Figure 5b), which is comparable to a value of $\sim$2800~km s$^{-1}$ obtained by \cite{mi18}, using only LASCO data points to compute this speed. The maximum CME acceleration is 5900~m s$^{-2}$. 

These properties place the 10 September 2017 CME in the top CMEs observed with the highest velocity, highest maximum acceleration and shortest acceleration time \cite{yu05, be11, liu13}. The flank speed of the CME was fitted using equation (3) at the southern flank, while the northern flank that shows slower expansion was fitted with a velocity function containing a simpler exponential acceleration component from \cite{ga03} (Supplementary Figure 4). }

\subsection*{Estimating Source Centroids and Uncertainties in Position}
{The tied-array beam pattern used in our observations is dense providing good coverage of the Sun within the tied-array beams field-of-view (Supplementary Figure 1). Due to the large number of beams covering the radio source contours, it is possible to determine centroid positions with a high accuracy of approximately $\pm$100~arcsec. The centroids of the sources were determined by fitting a 2D elliptical Gaussian to the source contours, similar to the analysis done by \cite{ko17}. Each radio source, even at 70\% to maximum intensity levels, has a large size that is covered by at least 8 tied-array beams, which allows us to determine the centroid position with high accuracy using the Gaussian fit method and, at the same time, calculate the uncertainties in centroid position. The intensity of the radio source, $I(x,y)$, in normalised arbitrary units, can be fitted by a 2D elliptical Gaussian using the following relation:
\begin{equation}
I (x,y) = A~exp( -a(x-x_0)^2 + 2b(x-x_0)(y-y_0) + c(y- y0)^2 )~,
\end{equation}
where $A$ is the amplitude, $x_0, y_0$ are the centroid positions and the coefficients $a$, $b$ and $c$ are given in terms of the rotation angle of the ellipse, $\theta$, and the rms lengths $\sigma_x$ and $\sigma_y$:
\begin{equation}
a = \frac{\cos(\theta)^2}{2\sigma_x^2} + \frac{\sin(\theta)^2}{2\sigma_y^2}~,
\end{equation}
\begin{equation}
b = -\frac{\sin(2\theta)}{4\sigma_x^2} + \frac{\sin(2\theta)}{4\sigma_y^2}~,
\end{equation}
\begin{equation}
c = \frac{\sin(\theta)^2}{2\sigma_x^2} + \frac{\cos(\theta)^2}{2\sigma_y^2}~.
\end{equation}

The centroid errors, $\delta x_0$ and $\delta y_0$, can then be estimated using the following relations \cite{ko17, co97}:  
\begin{equation}
\delta x_0 \approx \sqrt{\frac{2}{\pi}} \frac{\sigma_x}{\sigma_y} \frac{\delta I}{A} \theta_r
\end{equation}
\begin{equation}
\delta y_0 \approx \sqrt{\frac{2}{\pi}} \frac{\sigma_y}{\sigma_x} \frac{\delta I}{A} \theta_r
\end{equation}
where $\theta_r$ is the angular resolution. Using the above equations, we produced the centroids source positions shown in Figure 5a and Supplementary Figure 5. The errors in source position due to the Gaussian fit to the source are shown for the uncorrected sources (as triangles in Supplementary Figure 5). All other position error bars also contain a contribution from errors due to correcting for ionospheric shifts, discussed in the following section.
}

\subsection*{Ionospheric Corrections of Low Frequency Radio Sources}

{The ray refraction theory \cite{ko60} predicts a variation in radio source positions when rays travel through a plasma. The shift in source position varies with $f^{-2}$, where $f$ is the frequency of the radio source. The effect is expected to be significant at low radio frequencies for emission travelling through the ionosphere and this effect is termed ionospheric refraction. Solar radio bursts positions observed by LOFAR in the LBA band (10--90 MHz) are therefore affected by ionospheric refraction. In the case of solar observations, the displacement of the source is always negative towards zenith and for large zenith angles ($Z$) it is most affected by the spherical component of the ionosphere, in particular inside the F-layer \cite{st82}. The displacement, $\Delta\theta$, can be calculated according to the following relation:
\begin{equation}
\Delta\theta = -0.5 \frac{d}{R} \frac{f_p^2}{f^2}\tan Z \sec^2 Z ,
\end{equation}
where $d$ is the thickness of the F-layer, $R$ is the radius of curvature of the ionosphere and $f_p$ is the plasma frequency of the ionosphere.  }

{At the time of the LOFAR observation, the elevation of the Sun in Exloo Netherlands was low ($\sim 20$~degrees), therefore the zenith angle was large and the radio sources were most likely affected by ionospheric refraction. This was also indicated by the very high altitude of some of the radio sources (Group 3 at $>3~R_\odot$), which is in excess of altitude values predicted by electron density models \cite{zu14}, while the radio sources in Groups 1 and 2 appeared superimposed on the CME core, which was unexpected. It was necessary to consider ionospheric corrections of the raw radio images due to the low elevation of the Sun and the inconsistencies in positions of the observed radio sources. }

{The plasma frequency of the ionosphere varies depending on the stage of the solar cycle and for a most accurate estimation of $f_p$ we have looked at ionosonde data from the Global Ionospheric Radio Observatory \cite{giro} and computed an average of the peak $f_p$ from nearby stations at Juliusruh (Germany), Dourbes (Belgium) and Chilton (UK). The plasma frequency of the ionosphere at the time was low, $f_p=5\pm0.1$~MHz, since the Sun is in the declining phase of its cycle, meaning the displacements are less severe. The thickness of the F-layer was estimated to be $d = 200\pm50$~km and we attributed a large error to this quantity as it is highly dependent on the current ionospheric conditions and it is not obtainable from ionosonde data. The radius of curvature of the ionosphere represents the radius to the highest peak in density in the F2-layer (located at a height of $224\pm30$ km from the Earth's surface based on ionosonde data) and this quantity is $R = 6595\pm30$~km. Using equation (11), we obtained the shifts in source position due to the ionosphere. The deviations obtained at a frequency of 45~MHz were -$0.2\pm0.05$~$R_\odot$ in the x-direction and -$0.9\pm0.2$~$R_\odot$ in the y-direction. These corrections have been applied to all images of radio sources. It is necessary to note that these values represent the maximum deviations expected, as they were estimated using the peak plasma frequency of the ionosphere which does not cover the entire thickness of the F-layer. The main contribution to the errors obtained in the position shift is from the ratio $d/R$. It is also important to note that the ionospheric corrections do not affect the kinematics of the radio sources at a single frequency, as these corrections are frequency dependant.  The uncorrected centroids (triangles) and corrected centroids (filled circles) of herringbone groups 1, 2 and 3 are shown Supplementary Figure 5 for comparison. Ionospheric corrections are important when analyzing low frequency radio bursts positions and have been applied in our analysis of source locations. }

\section*{Data Availability}

The LOFAR dataset used was obtained under the project code DDT8\_005 and it is available in the LOFAR Long Term Archive (LTA; https://lta.lofar.eu). The I-LOFAR data can be obtained from http://data.lofar.ie or on request to observer@lofar.ie. The AIA and LASCO datasets are both available from the Virtual Solar Observatory project (http://vso.nso.edu). The SUVI data were made available by NOAA's National Centers for Environmental Information SUVI team as a community service to the solar physics community studying the September 2017 flaring events (http://doi.org/10.7289/V5FT8J93). These datatsets are also available from the authors upon request.

\begin{figure*}[ht]
\centering
\includegraphics[angle = 0, width = 400px, trim = 0px 0px 0px 0px ]{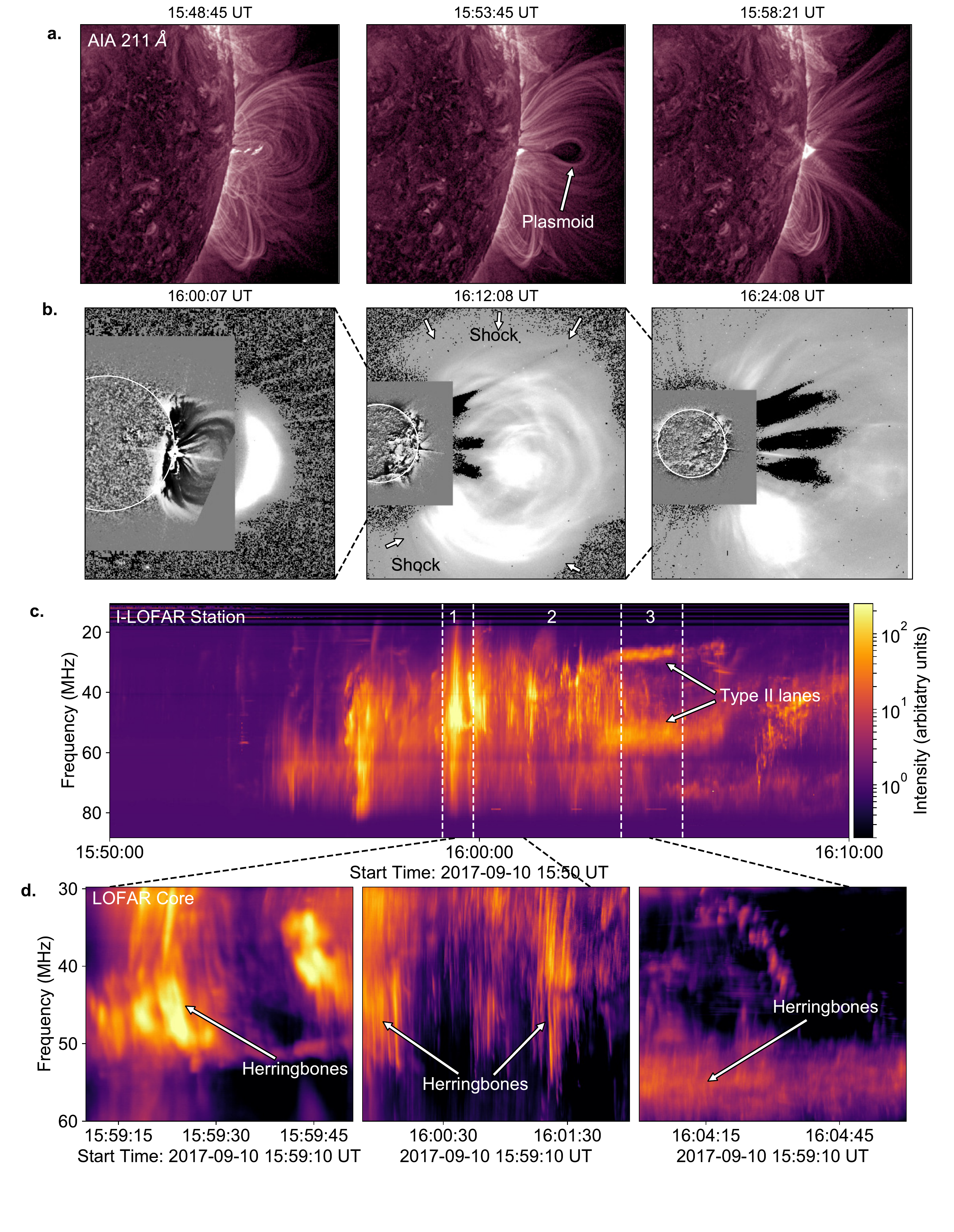}
\caption{The solar flare, CME and associated radio emission observed on 10 September 2017. (a) AIA 211~\AA~images showing the evolution of the plasmoid eruption that accompanied the 10 September 2017 X8.2 flare. (b) SUVI and LASCO composite images showing the expansion of the CME in the high corona. The white arrows denote the CME shock and the white circle denotes the solar limb. (c) I-LOFAR dynamic spectrum showing the radio signatures that accompanied the flare. (d) High-resolution LOFAR core dynamic spectra showing the fine structured radio shock signatures called herringbones. }
\end{figure*} 

\begin{figure*}[ht]
\centering
\includegraphics[angle = 0, width = 430px, trim = 100px 0px 0px 0px ]{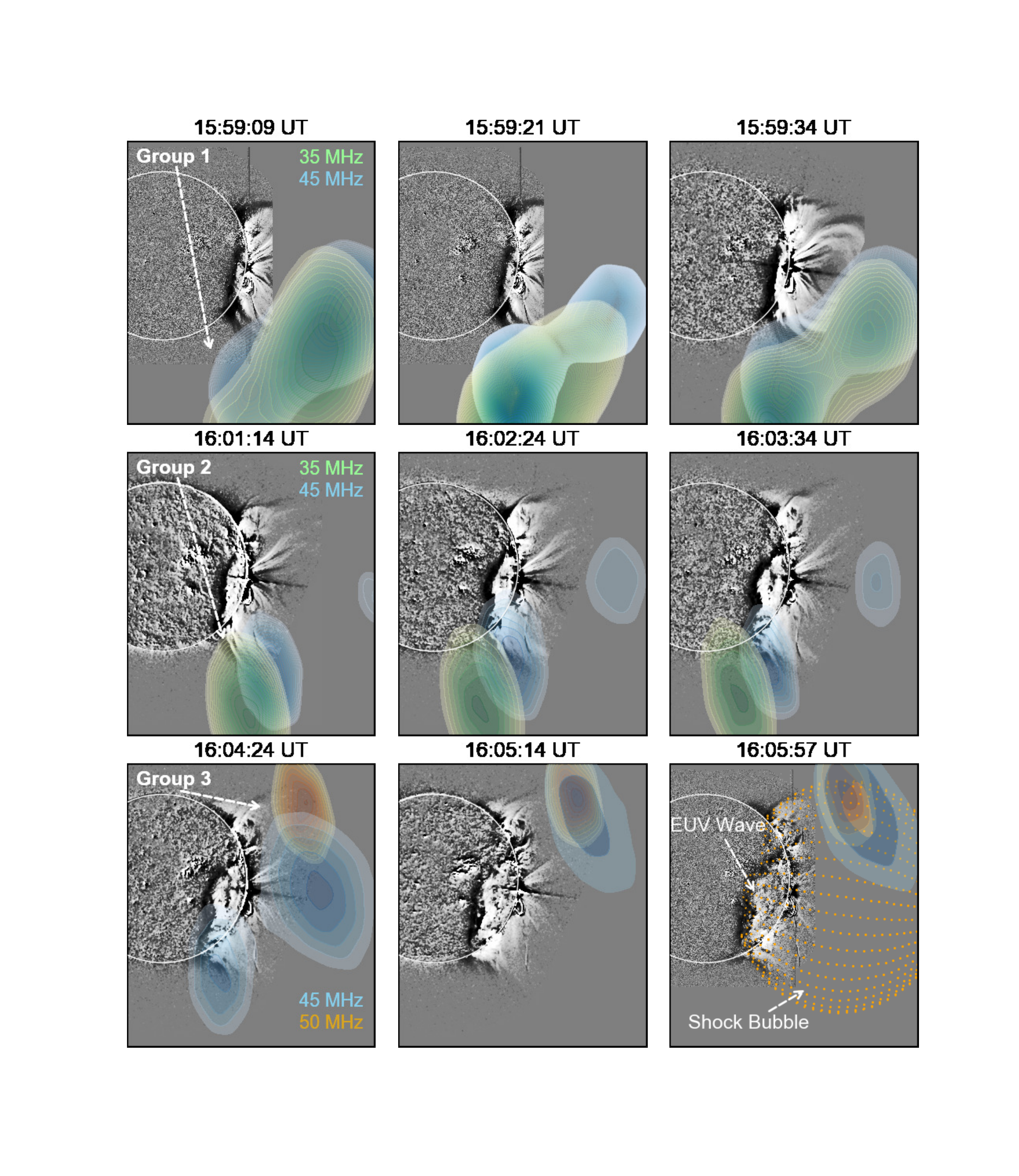}
\caption{Sequence of LOFAR tied-array filled contours showing the location of the radio shock signatures observed in Figure 1d. The top three panels show the Group 1 herringbones, middle panels show Group 2, and bottom panels show herringbones from Group 3. The LOFAR contours are overlaid on AIA 211~\AA~ and SUVI 193~\AA~ running difference images that show the CME eruption and the propagation of the coronal EUV wave. The contour levels are from 70\% to the maximum intensity for each image and the change in colour represents increasing intensity. The  green, blue and purple filled contours represent emission at 35, 45 and 50~MHz, respectively. The approximate 3D shock bubble surrounding the CME is overlaid in the last panel. The progression of the radio sources and propagation of the CME and EUV wave through time can also be seen in the Supplementary Movie, accompanying the paper. }
\end{figure*} 

\begin{figure*}[ht]
\centering
\includegraphics[angle = 0, width = 460px, trim = 100px 0px 0px 0px ]{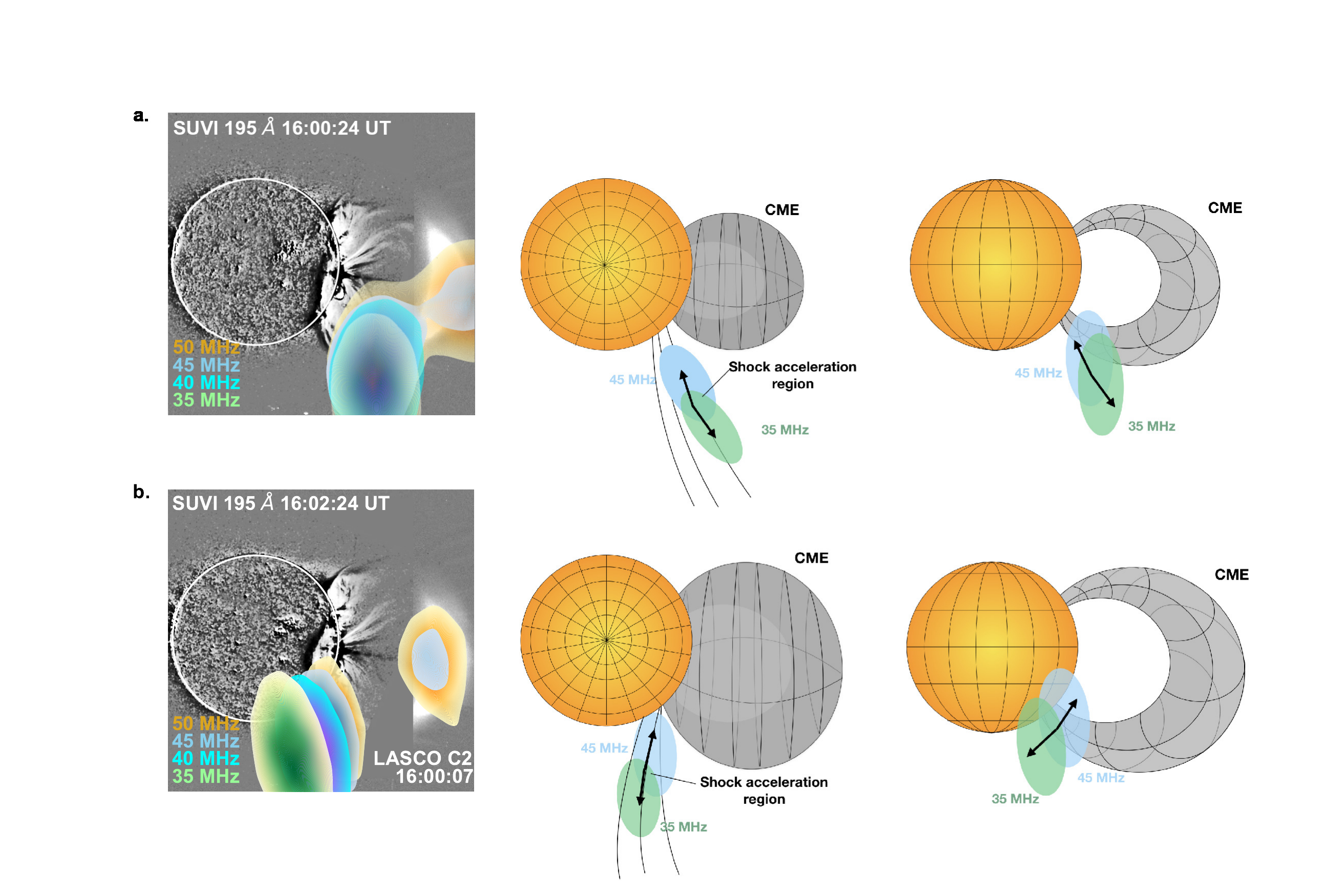}
\caption{ Location of herringbone sources at four frequencies from 35 to 50~MHz at 16:00:24~UT (a) and 16:02:24~UT (b). The middle panels show cartoons from a solar north pole perspective of the possible geometry of open magnetic field lines along which herringbone electrons escape to explain the source locations on the left. The right panels show cartoons of the same herringbone sources from the line-of-sight perspective.}
\end{figure*} 

\begin{figure*}[ht]
\centering
\includegraphics[angle = 0, width = 460px, trim = 70px 0px 0px 40px ]{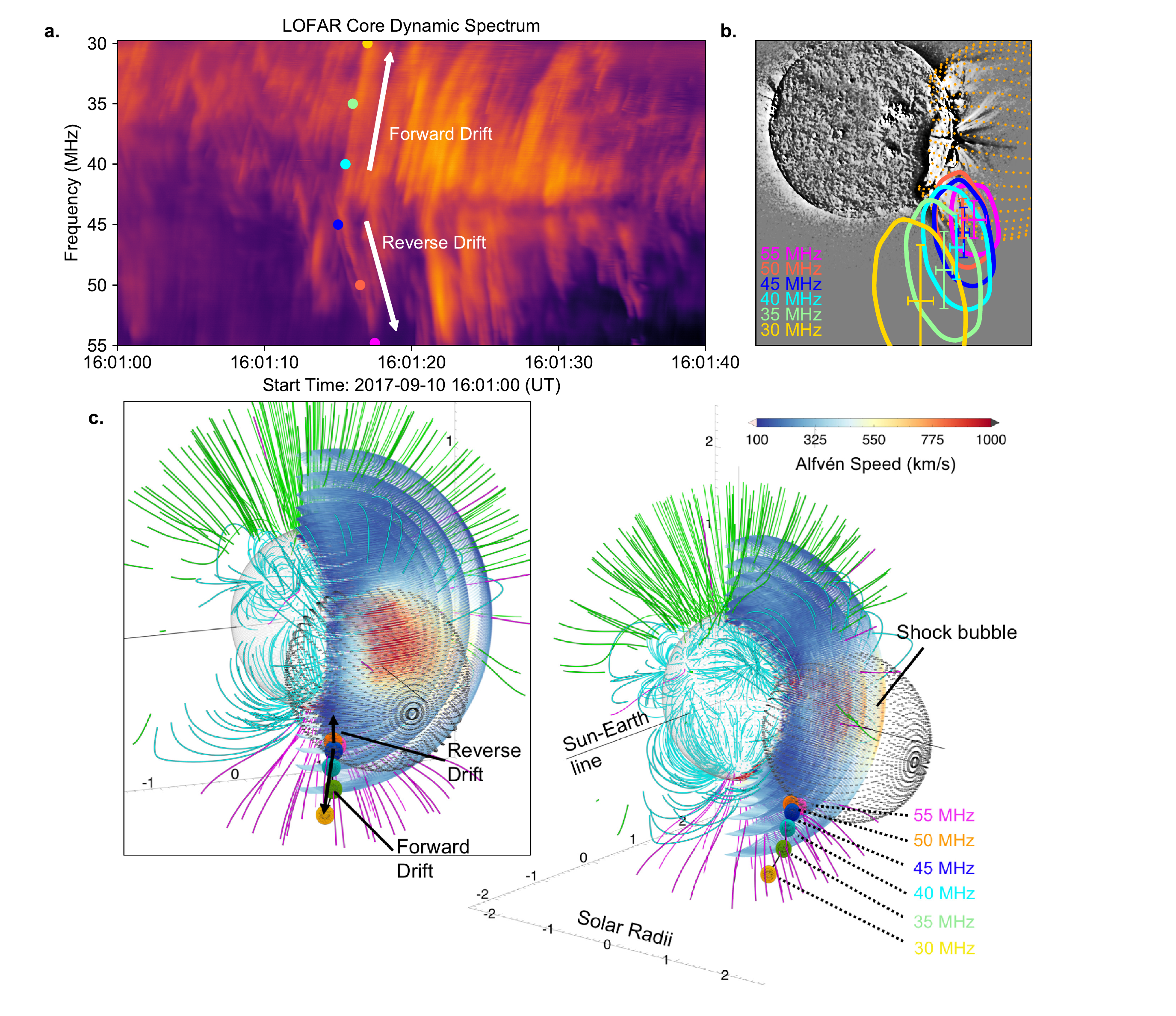}
\caption{The location in 2D and 3D of an individual herringbone. (a) Dynamic spectrum of herringbones in Group 2, where an individual herringbone is selected with forward and reverse drift components. (b) 70\% contours at six frequencies of the herringbone source location in (a) overlaid on the SUVI running difference image of the CME and an estimate of the 3D shock bubble surrounding the CME (orange dots). The error bars represent the uncertainty in determining the centroid position, as well as the errors in correcting for ionospheric refraction. (c) Model of the location of Alfv\'en speed surfaces in 3D space relative to the CME shock bubble (black dots), open magnetic field lines (purple and green lines) and the location of the de-projected radio source centroids in (b) at six frequencies. The herringbone sources are located on top of the shock bubble in areas of low Alfv\'en speed and open field lines. }
\end{figure*} 

\begin{figure*}[ht]
\centering
\includegraphics[angle = 0, width = 460px, trim = 70px 0px 0px 0px ]{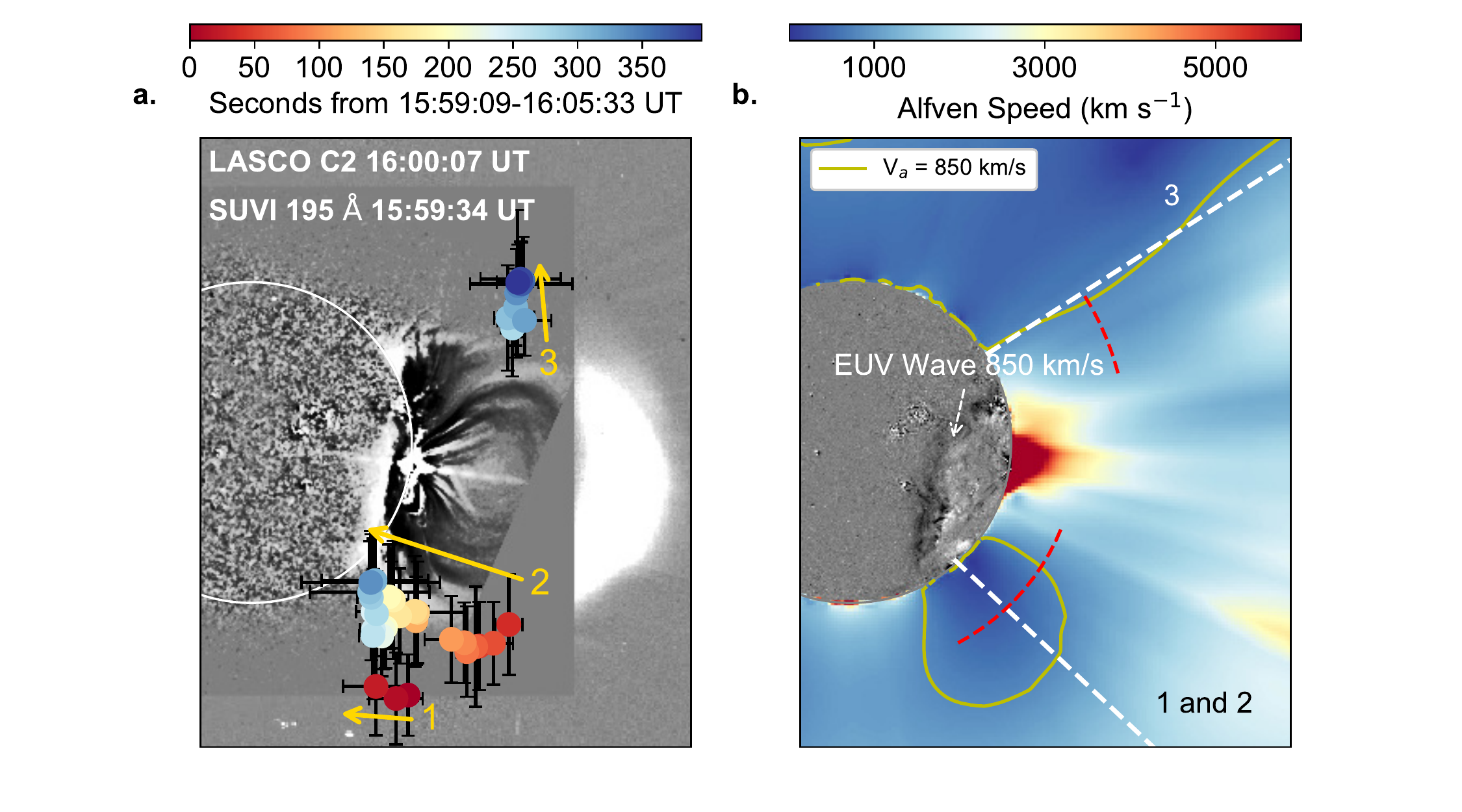}
\caption{The movement of radio sources through time and the Alfv\'en speed environment. (a) Composite image showing the full extent of the CME in the low corona composed of SUVI 193~\AA~ running difference image at 15:59:34~UT and LASCO C2 running difference image at 16:00:07~UT. Overlaid are the centroids (filled circles) of the radio shock signatures colour-coded through time from 15:59:09~UT to 15:05:33~UT, showing the direction of the herringbone sources. The error bars represent the errors in determining the centroid position, as well as the errors in correcting for ionospheric refraction. (b) Composite image of AIA 211~\AA~ running difference image of the EUV wave at 16:05~UT and plane-of-sky Alfv\'en speed map of the solar corona. The labels 1, 2 and 3 accompanying the white dashed lines denote the radial directions that pass through the locations of Groups 1, 2 and 3. Groups 1, 2 and 3 occur in regions of low Alfv\'en speed in the plane-of-sky. The red dashed arcs in this figure are used to compute the CME flank velocity and the Alfv\'en speed along the expanding flanks direction (see Figure 6). }
\end{figure*}

\begin{figure*}[ht]
\centering
\includegraphics[angle = 0, width = 450px, trim = 40px 0px 0px 0px ]{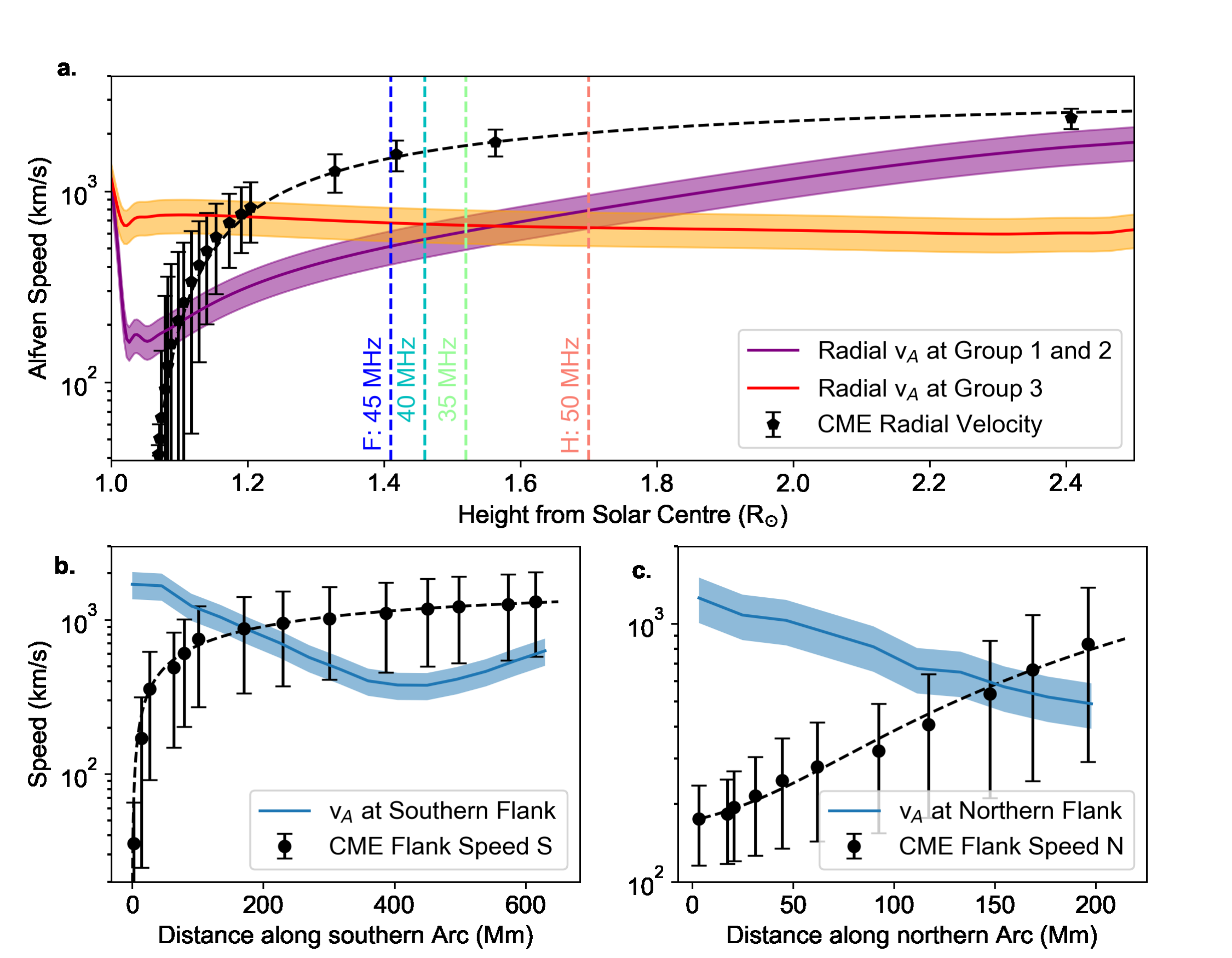}
\caption{The CME and Alfv\'en speeds. (a) CME speed as a function of height from the photosphere (1~R$_\odot$) computed using AIA, SUVI and LASCO images. Overlaid is the Alfv\'en speed along the radial profiles passing through the location of Group 1 and 2 (orange) and Group 3 (green) in Figure 5b. The heights of herringbones at the fundamental frequency of 35 (light green), 40 (cyan) and 45~MHz (purple) and harmonic frequency of 50~MHz (pink) are shown by the dashed vertical lines. (b), (c) CME speed (black) and Alfv\'en speed (blue) at the southern and northern flank, respectively, computed along the red dashed arcs in Figure 5b. The CME is super-alfv\'enic in the radial direction and at both flanks.}
\end{figure*}

\end{document}